%% file: sample61.tex
\documentclass[modern]{aastex61}

 \accepted{January 1, 2018}
\submitjournal{RNAAS}

\shorttitle{SAMPLING Data Release 1}
\shortauthors{Wang et al.}

\begin{document}

\title{First Data Release of the ESO-ARO Public Survey SAMPLING --- SMT ``All-sky'' Mapping of PLanck Interstellar Nebulae in the Galaxy}

\correspondingauthor{Ke Wang}
\email{kwang@eso.org}

\author[0000-0002-7237-3856]{Ke Wang}
\affil{European Southern Observatory,
Karl-Schwarzschild-Str. 2,
85748 Garching,
Germany}

\author{Sarolta Zahorecz}
\affil{National Astronomical Observatory of Japan, Mitaka, Tokyo 181-8588, Japan}
\affil{Osaka Prefecture University, Osaka 599-8531, Japan}

\author{Maria R. Cunningham}
\affil{School of Physics, University of New South Wales, Sydney, NSW 2052, Australia}

\author{L. Viktor T{\'o}th}
\affil{Department of Astronomy, E\"{o}tv\"{o}s Lor\'{a}nd University, Budapest, Hungary}

\author{Tie Liu}
\affil{East Asian Observatory, 660 N. A'ohoku Place, Hilo, HI 96720, USA}

\author{Xing Lu}
\affil{National Astronomical Observatory of Japan, Mitaka, Tokyo 181-8588, Japan}

\author{Yuan Wang}
\affil{Max-Planck Institute f\"{u}r Astronomie, K\"{o}nigstuhl 17, D-69117 Heidelberg, Germany}

\author{Giuliana Cosentino}
\affil{Department of Physics and Astronomy, University College London, London WC1E 6BT, UK}

\author{Ren-Shiang Sung}
\affil{National Tsing Hua University, Hsinchu, Taiwan}

\author{Vlas Sokolov}
\affil{Max Planck Institute for Extraterrestrial Physics, Giessenbachstrasse 1, 85748 Garching, Germany}


\author{Shen Wang}
\affil{National Astronomical Observatories, Beijing, China}

\author{Yuwei Wang}
\affil{School of Astronomy and Space Science, Nanjing University, Nanjing, China}

\author{Zhiyu Zhang}
\affil{European Southern Observatory,
Karl-Schwarzschild-Str. 2,
85748 Garching,
Germany}

\author{Di Li}
\affil{National Astronomical Observatories, Beijing, China}

\author{Kee-Tae Kim}
\affil{Korea Astronomy and Space Science Institute, Daejeon, Republic of Korea}

\author{Ken'ichi Tatematsu}
\affil{National Astronomical Observatory of Japan, Mitaka, Tokyo 181-8588, Japan}

\author{Leonardo Testi}
\affil{European Southern Observatory,
Karl-Schwarzschild-Str. 2,
85748 Garching,
Germany}

\author{Yuefang Wu}
\affil{Department of Astronomy, Peking University, Beijing, China}

\author{Ji Yang}
\affil{Purple Mountain Observatory, Nanjing, China}

\author{SAMPLING Collaboration}




\input{mycmd.tex}

\section{Introduction} \label{sec:intro}

In the last decades, a large effort has been made to characterize properties of star formation regions in the Galaxy through observations in both dust continuum and spectral lines. However, most of the previous studies have concentrated on the inner Galactic plane. Consequently, our knowledge of star formation in the outer Galaxy and high latitudes is so far only scarcely constrained.
This lack of knowledge has posed a long-standing important question: how does star formation (or even cloud evolution) depend on the Galactic environment (metalicity, pressure, turbulence, shearing, etc.)?
The recently released \textit{Planck} Catalog of Galactic Cold Clumps (PGCCs, \citealt{Planck2015-PGCC})
provides the first unbiased all-sky catalog of intrinsically cold clumps with a resolution of $\sim 5'$ (Figure 1). Many PGCC sources, especially those outside the Galactic plane, are new and have never been explored before.

\section{The SAMPLING Survey}
SAMPLING, the SMT ``All-sky'' Mapping of PLanck Interstellar Nebulae in the Galaxy,\footnote{Project website: \url{http://sky-sampling.github.io}}
is an ongoing ESO-ARO Public Spectroscopic Survey\footnote{\url{http://www.eso.org/sci/activities/call_for_public_surveys.html}}
aiming to obtain \coa\ and \cob\ (2--1) spectral-imaging of about 600 highly reliable PGCCs, using the 10-m Heinrich Hertz Submillimeter Telescope (SMT) of Arizona Radio Observatory (ARO). The sources were selected from an earlier release of PGCCs with high signal-to-noise ratios and low dust temperature ($T_{\rm dust} < 14$\,K, \citealt{Planck2011-ECC}), and have been detected in \cob (1--0) \citep[Figure 1;][]{WuYF2012-Planck}.

SAMPLING has been awarded 618 hours of SMT time over 3 years (2015-2018).
Observations with the SMT are being conducted in on-the-fly mode with over Nyquist sampling, usually on a $5'\times5'$ area for each source. The data is calibrated in GILDAS\footnote{\url{http://www.iram.fr/IRAMFR/GILDAS}} after flagging bad channels. The final data cubes are in units of main beam brightness temperature $T_{\rm mb}$ (therefore directly comparable to other observations and simulations), adopting a main beam efficiency of 0.7. The cubes are gridded in $8''$ pixels, with an effective resolution of $36''$. The channel width is 0.33\kms\ and the RMS noise is required to be $T_{\rm mb}<0.2$ K in order to pass our data quality assurance.

\section{Data Release}
Here, we make the first data release (DR1)\footnote{\url{http://dx.doi.org/10.7910/DVN/0L8NHX}}
of the SAMPLING survey. DR1 comprises of 124 fields distributed in $70^\circ < l < 216^\circ$, $-35^\circ < b < 25^\circ$,
observed from 2015 November 27 to 2016 February 23 and from 2016 December 3 to 24. We provide FITS cubes, source list, and sky coverage. Figure 1 right panel shows an example source demonstrating the potential of the data.

The SAMPLING data resolves dense molecular clouds and enables revealing of clump structure, turbulence, kinematics, and star formation (e.g., high-velocity outflows). Most of the SAMPLING sources have never been explored at sub-clump resolution.
Once completed, SAMPLING and complementary surveys \citep[e.g.,][]{LiuT17} will initiate the first major step forward to characterize molecular clouds and star formation on truly Galactic scales.

The community is encouraged to download and use the data following the standard acknowledgments in the DR1 link. Collaboration with the SAMPLING team is particularly encouraged.
Several papers have made use of the SAMPLING data \citep{Juvela17,LiuT17,YuanJH17-HMSC}.
It is worth noting that, the data is not only useful for Galactic star formation; for example, it can be used to evaluate the ``contamination'' of Galactic CO emission, when concerned \citep{Khatri2015-SZ}.
We plan to release the full data in 2018 along with a more detailed paper, when all the data has been taken.




\begin{figure}
\centering
\includegraphics[trim=2cm 4cm 3cm 4cm, clip, width=0.68\textwidth,angle=0]{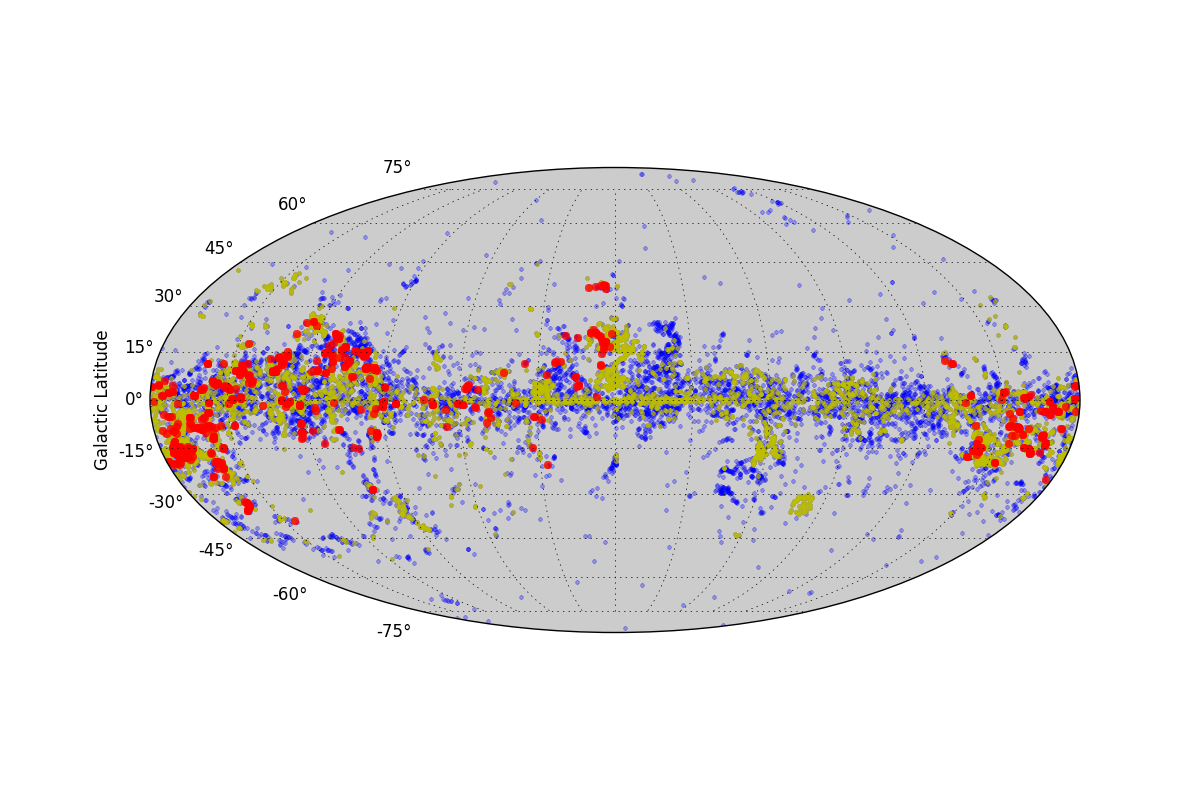}
\includegraphics[trim=1cm 1cm .5cm 2cm, clip, width=0.3\textwidth,angle=0]{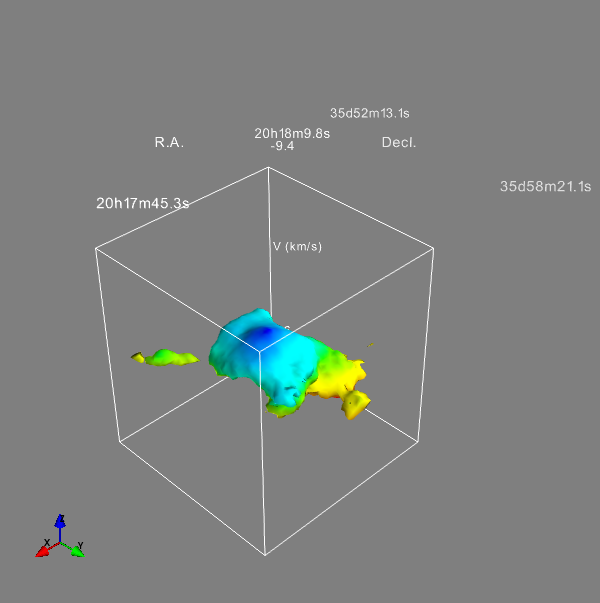}
\caption{
\textbf{Left}:
All-sky distribution of the PGCC sources (13188, blue), with estimated distance (4655, yellow), and mapped by Purple Mountain Observatory 14-m telescope (618, red) in \coa/\cob/\coc\ (1--0). The red coded sources are targets for the SAMPLING survey in \coa/\cob\ (2--1) with SMT.
\textbf{Right}: \cob\ (2--1) cube of an example SAMPLING source. The 3D view shows RA, Dec, and velocity axes, with blue-/red-shifted radial velocity color coded in blue/red. The cube reveals multiple radial filaments feeding the central, star-forming hub (Harju, J. et al., in prep).
}
\label{fig:SAMPLING}
\end{figure}

\acknowledgments
The data presented in this paper are based on the ESO-ARO programme ID 196.C-0999(A).

%

\vspace{5mm}
\facilities{SMT}



\bibliographystyle{aasjournal}
\bibliography{my.astro}



\end{document}

%% file: mycmd.tex
\newcommand{\skipthis}[1]{}
\newcommand{\hii}{{\rm H}{\sc ii}}
\newcommand{\uchii}{{\rm UCH}{\sc ii}}
\newcommand{\lsim}{${\raisebox{-.9ex}{$\stackrel{\textstyle<}{\sim}$}}$ }
\newcommand{\gsim}{${\raisebox{-.9ex}{$\stackrel{\textstyle>}{\sim}$}}$ }
\newcommand{\mycomment}[1]{[{\bf comment:} #1]}

\newcommand{\coa}{$^{12}$CO}
\newcommand{\cob}{$^{13}$CO}
\newcommand{\coc}{C$^{18}$O}
\newcommand{\eup}{$E_{\rm up}$}
\def\Ncol{$N_{\rm H_2}$}

\def\nh2d{$\rm{NH_2D}$}
\def\nhd{$\rm{NH_2D}$}

\def\nh3{$\rm{NH_3}$}
\def\NH3{$\rm{NH_3}$}
\def\am{$\rm{NH_3}$}

\def\n2hp{$\rm{N_2H^+}$}
\def\NTH{$\rm{N_2H^+}$}
\def\nhp{$\rm{N_2H^+}$}
\def\ndp{$\rm{N_2D^+}$}

\def\h2o{$\rm{H_2O}$}
\def\h2{$\rm{H_2}$}
\def\water{H$_2$O}

\def\meth{CH$_3$OH}
\def\hcop{$\rm{HCO^+}$}

\def\msolar{\,$M_\odot$}
\def\msun{\,$M_\odot$}
\def\Mout{\,$\dot{M}_{\rm out}$}
\def\Macc{\,$\dot{M}_{\rm acc}$}
\newcommand{\mj}{$M_{\rm J}$}
\newcommand{\mturb}{$M_{\rm turb}$}
\def\lsolar{\,$L_\odot$}
\def\lsun{\,$L_\odot$}

\def\um{\,$\mu\mathrm{m}$}
\def\mm{\,mm}
\def\kms{\,km~s$^{-1}$}
\def\cm2{\,$\rm{cm^{-2}}$}
\def\cm3{\,$\rm{cm^{-3}}$}
\def\cms{\,$\rm{cm^{-2}}$}
\def\cmc{\,$\rm{cm^{-3}}$}

\newcommand{\ghz}{\,GHz}
\newcommand{\mhz}{\,MHz}
\newcommand{\khz}{\,kHz}
\newcommand{\jy}{\,Jy}
\newcommand{\jyb}{\,Jy\,beam$^{-1}$}
\newcommand{\mjy}{\,mJy}
\newcommand{\mjyb}{\,mJy\,beam$^{-1}$}

\newcommand{\tsys}{$T_{\rm sys}$}

\def\dv{$\Delta V$}
\def\vlsr{$v\rm{_{LSR}}$}
\def\vmax{$v_{\rm max}$}
\def\vmin{$v_{\rm min}$}
\def\tdyn{$t_{\rm dyn}$}
\def\tff{$t_{\rm ff}$}
\def\tacc{$t_{\rm acc}$}
\def\tkh{$t_{\rm KH}$}

\newcommand{\sigth}{$\sigma_{\rm Therm}$}
\newcommand{\signt}{$\sigma_{\rm NT}$}

\def\jk{$(J,K)$}
\def\11{(1,1)}
\def\22{(2,2)}
\def\33{(3,3)}
\def\44{(4,4)}
\def\55{(5,5)}
\def\66{(6,6)}
\def\t21{$T_{21}$}
\def\r31{$R_{31}$}
\def\tba{$T_{21}$}
\def\rca{$R_{31}$}
\newcommand{\trot}{$T_{\rm rot}$}
\newcommand{\tkin}{$T_{\rm kin}$}
\newcommand{\tex}{$T_{\rm ex}$}
\newcommand{\tbg}{$T_{\rm bg}$}
\newcommand{\Tex}{$T_{\rm ex}$}
\newcommand{\Tmb}{$T_{\rm mb}$}

\newcommand{\note}[1]{\textcolor{red}{[{\bf Ke's note: #1}]}}

\def\uv{$(u,v)$}
\def\xy{$(x,y)$}

\def\ga{G28.34+0.06}
\def\gapa{G28.34-P1}
\def\gapb{G28.34-P2}
\def\gb{G11.11{\textendash}0.12}
\def\gbpa{G11.11-P1}
\def\gbpb{G11.11-P6}
\def\gc{G30.88+0.13}
\def\gcca{G30.88-C1}
\def\gccb{G30.88-C2}
\def\gd{G10.6{\textendash}0.4}

\def\apex{Atacama Pathfinder EXperiment} 
\def\alma{Atacama Large Millimeter/submillimeter Array}
\def\atca{Australia Telescope Compact Array}
\def\vla{Very Large Array}
\def\evla{Expanded Very Large Array}
\def\sma{Submillimeter Array}
\def\carma{Combined Array for Research in Millimeter Astronomy}
\def\gbt{Green Bank 100m Telescope}
\def\cso{Caltech Submillimeter Observatory}
\def\pmo{Purple Mountain Observatory}
\def\kosma{K\"{o}lner Observatorium f\"{u}r SubMillimeter Astronomie}
\def\iram{Institut de Radioastronomie Millim\'{e}trique}
\def\jcmt{James Clerk Maxwell Telescope}

\def\iras{\emph{Infrared Astronomical Satellite}}
\def\iso{\emph{Infrared Space Observatory}}
\def\msx{\emph{Midcourse Space Experiment}}

\def\spt{\emph{Spitzer}}
\def\her{\emph{Herschel}}
\def\plk{\emph{Planck}}

\newcommand{\aips}{Astronomical Image Processing System}
\newcommand{\casa}{Common Astronomy Software Applications}
\newcommand{\mir}{Millimeter Interferometry Reduction}
\newcommand{\miriad}{Multichannel Image Reconstruction, Image Analysis and Display}
\newcommand{\gildas}{Grenoble Image and Line Data Analysis Software}

\newcommand{\aipsurl}{\url{ http://www.aips.nrao.edu}}
\newcommand{\casaurl}{\url{ http://casa.nrao.edu}}
\newcommand{\gbtidlurl}{\url{ http://gbtidl.nrao.edu}}
\newcommand{\mirurl}{\url{ http://www.cfa.harvard.edu/~cqi/mircook.html}}
\newcommand{\miriadurl}{\url{ http://www.cfa.harvard.edu/sma/miriad}}

\newcommand{\aksma}{The Submillimeter Array is a joint project between the Smithsonian Astrophysical Observatory
and the Academia Sinica Institute of Astronomy and Astrophysics and is funded by the
Smithsonian Institution and the Academia Sinica.}

\newcommand{\akvla}{The National Radio Astronomy Observatory
is a facility of the
National Science Foundation operated under cooperative agreement by Associated Universities, Inc.}
\let \akevla = \akvla
\let \aknrao = \akvla

\newcommand{\akkosma}{The KOSMA 3\,m radiotelescope at
Gornergrat-S\"ud Observatory is operated by the University of
Cologne and supported by special funding from the Land NRW. The
Observatory is administered by the Internationale Stiftung
Hochalpine Forschungsstationen Jungfraujoch und Gornergrat,
Bern.}

\let \ndash = \textendash
\let \amp = \&